\documentclass{article}
\usepackage{spconf,graphicx}
\usepackage{amsmath,amssymb}
\usepackage{hyperref}
\usepackage{xcolor}
\usepackage[ruled,linesnumbered]{algorithm2e}
\usepackage{multicol}
\usepackage{cuted}
\usepackage[T1]{fontenc}
 
	\usepackage{pgfplots}
  \pgfplotsset{compat=newest}
\newlength\figurewidth

\def\RR{\mathbb{R}}

\def\CC{\mathbb{C}}

\def\x{\vect{x}}
\def\u{\vect{u}}
\def\vv{\vect{v}}
\def\y{\vect{y}}
\def\z{\vect{z}}

\def\Mr{M_\text{R}}
\def\Mh{M_\text{H}}
\def\Ml{M_\text{L}}

\newcommand{\norm}[1]{\|#1\|}

\newcommand{\vect}[1]{\mathbf{#1}} 

\newcommand{\argmin}{\mathop{\operatorname{arg~min}}}
\newcommand{\modulo}{\mathop{\operatorname{mod}}}

\newcommand{\sgn}{\mathrm{sgn}}

\newcommand{\tc}{\ensuremath{\theta_\text{c}}}

\newcommand{\dsdr}{\ensuremath{\mathrm{\Delta SDR}}}

\newcommand{\qm}[1]{``#1''}  

\newcommand{\igg}{\iota_{\Gamma}(\x)}
\newcommand{\il}{\iota_{\ell_0 \leq k}(\z)}
\newcommand{\uv}[1]{``#1''}

\title{A proper version of synthesis-based Sparse Audio Declipper}
\name{Pavel Z\'avi\v{s}ka$^{\star}$ \qquad 
Pavel Rajmic$^{\star}$ \qquad 
Ond\v{r}ej Mokr\'y$^{\star}$ \qquad 
Zden\v{e}k Pr\r{u}\v{s}a$^{\dagger}$
\thanks{The authors thank S.\,Kiti\'{c} for providing them with his implementation of the SPADE algorithms and for discussion.
The work was supported by the joint project of the FWF and the Czech Science Foundation (GA\v{C}R): numbers I 3067-N30 and 17-33798L, respectively. 
Research described in this paper was financed by the National Sustainability Program under grant LO1401. 
Infrastructure of the SIX Center was used.} 
}	

\address{$^{\star}$Signal Processing Laboratory, Brno University of Technology, Brno, Czech Republic \\
$^{\dagger}$Acoustics Research Institute, Austrian Academy of Science, Vienna, Austria \\
Email: $^{\star}$\{xzavis01, rajmic, 170583\}@vutbr.cz, $^{\dagger}$zdenek.prusa@oeaw.ac.at \\
} 

\begin{document}
\maketitle
\begin{abstract} 
Methods based on sparse representation have found great use in the recovery of audio signals degraded by clipping.
The state of the art in declipping within the sparsity-based approaches has been achieved by the SPADE algorithm by Kiti\'c et.\,al.\ (LVA/ICA'15).
Our recent study (LVA/ICA'18) has shown that although the original \mbox{S-SPADE} can be improved 
such that it converges faster than the \mbox{A-SPADE}, the restoration quality is significantly worse.
In the present paper, we propose a new version of S-SPADE.
Experiments show that the novel version of S-SPADE outperforms its old version in terms of restoration quality, and that it is comparable with the \mbox{A-SPADE}
while being even slightly faster than A-SPADE.
\end{abstract}

\vspace{-.5em}
\section{Introduction}
\vspace{-.2em}

Clipping is one of the common types of signal degradation.
It is usually caused by an element in the signal path whose dynamic range is insufficient compared to the dynamics of the signal.
This fact causes the peaks of the signal to be cut (saturated).
More exactly, in the so-called \emph{hard clipping}, samples of the input signal \(\x \in \RR^N\) that exceed the dynamic range given by the thresholds \([-\tc, \tc]\)
are modified such that the signal output can be described by the formula
\begin{equation}
\y [n] = \left\{
\begin{aligned}
&\x [n] &\text{for} \hspace{1em} &|\x[n]| < \tc, \\
&\tc \cdot \sgn(\x[n]) &\text{for} \hspace{1em}  &|\x[n]| \geq \tc.
\end{aligned}
\right.
\label{eq:clipping}
\end{equation}

Due to the great number of higher harmonics that appear in
the clipped signal, the clipping has a negative effect on the perceived audio quality \cite{Tan2003}.
Therefore it is inevitable to perform restoration, so-called \emph{declipping}, i.e.\ a~recovery of the clipped samples, based on the observed signal $\y$.

In the hard clipping case, which is the context of the paper, the signal samples can be divided into three sets \(R\), \(H\), and \(L\), which correspond to \uv{reliable} samples and samples that have been clipped to \uv{high} and \uv{low} clipping thresholds, respectively.
To select only samples from a specific set, the \emph{restriction operators} \(\Mr, \Mh, \Ml\) will be used.
These operators can be understood as linear projectors or identity matrices with specific columns removed that correspond to the three cases.

With the additional information that the positive clipped samples should lie above the \(\tc>0\)
and the negative clipped samples below \(-\tc\), a~set of feasible solutions \(\Gamma\) is defined
as a (convex) set of time-domain signals such that 
\begin{equation}
	\label{eq:gamma}
	\Gamma \hspace{-1.5pt}=\hspace{-1.5pt} \Gamma(\y) 
	\hspace{-1.5pt}=\hspace{-1.5pt} \{\tilde{\x} 
	\hspace{0.25em}|\hspace{0.25em} 
	\Mr\tilde{\x} \hspace{-1.5pt}  =   \hspace{-1.5pt} \Mr\y, 
	\Mh\tilde{\x} \hspace{-1.5pt}\geq  \hspace{-1.5pt} \tc, 
	\Ml\tilde{\x} \hspace{-1.5pt}\leq  \hspace{-1.5pt} -\tc 
	\}.
\end{equation}
The restored signal, \(\hat{\x}\), is naturally required to be a~member of the set, i.e.\ \(\hat{\x} \in \Gamma\).
Finding \(\hat{\x}\)
is an ill-posed problem since there are infinitely many possible solutions. 
A possible way to treat this problem is to exploit the fact that audio signals are sparse with respect to a (time-)frequency transform.
In other words, the goal is to find the signal \(\hat{\x}\) from the set \(\Gamma\) of the highest sparsity.

In the past, several approaches to declipping were introduced. 
Focusing on the sparsity-based methods, the very first method using the sparsity assumption was reported in \cite{Adler2011:Declipping};
it was based on the greedy approximation of a~signal within the reliable parts. 
In \cite{Weinstein2011:DeclippingSparseland}, convex optimization was used.
According to \cite{SiedenburgKowalskiDoerfler2014:Audio.declip.social.sparsity}, adding the structure to the coefficients may lead to the improvement in the restoration quality.
The authors of \cite{Kitic2013:Consistent.iter.hard.thresholding} used an iterative hard thresholding algorithm that was constrained to solve the declipping task and in \cite{KiticBertinGribonval2014:AudioDeclippingCosparseHardThresholding} reformulated the task to the analysis approach to the sparsity.

On top of these approaches, non-negative matrix factorization has also been recently adopted to audio declipping \cite{BilenOzerovPerez2015:declipping.via.NMF}.

As far as the authors know, \cite{Kitic2015:Sparsity.cosparsity.declipping} presented the current state-of-the-art,
a~heuristic
declipping algorithm for both the analysis and the synthesis models (the SPADE algorithm). 
Until recently, the synthesis variant was considered significantly slower due to the difficult projection step, but \cite{ZaviskaRajmicPrusaVesely2018:RevisitingSSPADE} has shown that the opposite is true---the acceleration makes the synthesis model require even fewer iterations to converge, making it faster than the analysis variant.
Unfortunately, the restoration quality of the synthesis variant has been  shown to be substantially worse. 
In this paper, the problem of the original synthesis variant is briefly explained and a new, more proper, synthesis version of the algorithm is presented.
%

\section{SPADE algorithms}
	
SPADE (SParse Audio DEclipper) \cite{Kitic2015:Sparsity.cosparsity.declipping} by Kiti\'c et.\,al.\ is a~sparsity-based heuristic declipping algorithm.
It is derived using the Alternating Direction Method of Multipliers (ADMM),
which is briefly revised first.
For details and proofs, see
\cite{ZaviskaMokryRajmic2018:SPADE_DetailedStudy}.

	\subsection{ADMM}
The ADMM \cite{Boyd2011ADMM} is a means for solving problems of the form
\(\min f(\x) + g(A\x)\), or equivalently
\begin{equation}
	\min_{\x,\z} f(\x) + g(\z) \hspace{1em} \text{s.t.} \hspace{1em} A\x - \z = 0,
	\label{eq:problem_formulation}
\end{equation}
where \(\x \in \mathbb{C}^N, \z \in \mathbb{C}^P\) and \(A : \mathbb{C}^N \rightarrow \mathbb{C}^P\) is a linear operator. 
ADMM is based on minimizing the Augmented Lagrangian, defined for \eqref{eq:problem_formulation} as:
\begin{equation}
	L_\rho(\x,\y,\z) = f(\x) + g(\z) + \y^{\top}(A\x-\z) + \frac{\rho}{2}\|A\x - \z\|^2_2,
	\label{eq:augmented_lagrangian}
\end{equation}
where \(\rho > 0\) is called the \emph{penalty parameter}.
The ADMM consists of three steps:
minimization of \eqref{eq:augmented_lagrangian} over \(\x\), over \(\z\), and the update of the dual variable, formally \cite{Boyd2011ADMM}: 
\begin{subequations}\label{eq:ADMM_unscaled}
	\begin{align}
		\x^{(i+1)} &= \argmin_\x L_\rho \left(\x, \z^{(i)}, \y^{(i)}\right), \\
		\z^{(i+1)} &= \argmin_\z L_\rho \left(\x^{(i+1)}, \z, \y^{(i)} \right), \\
		\y^{(i+1)} &= \y^{(i)} + \rho \left( A\x^{(i+1)} - \z^{(i+1)}\right).
	\end{align}
\end{subequations}
The ADMM can be often seen in the so-called \emph{scaled form}, which we obtain by substituting a dual variable \(\y\) with the scaled dual variable \(\u = \y/\rho\).

\subsection{SPADE}
\label{subsec:spade}
The SPADE algorithm \cite{Kitic2015:Sparsity.cosparsity.declipping} approximates the solution of the following NP-hard
regularized inverse problems
\begin{subequations}\label{eq:problem_const}
	\begin{align}
	\label{subeq:analysis}
	&\min_{\x,\z} \|\z\|_0 \ \, \text{s.t.}
	\ \,\x\in\Gamma(\y) \ \text{and}\ 
	\norm{A\x-\z}_2 \leq\epsilon, \\
	\label{subeq:synthesis}
	&\min_{\x,\z} \|\z\|_0 \ \, \text{s.t.}
	\ \,\x\in\Gamma(\y) \ \text{and}\ 	
	\norm{\x-D\z}_2 \leq\epsilon,
	\end{align}
\end{subequations}
where \eqref{subeq:analysis} and \eqref{subeq:synthesis} represent the problem formulation for the analysis and the synthesis variant, respectively.
Here, \(\Gamma\) denotes the set of feasible solutions (see Eq.\,\eqref{eq:gamma}),
\(\x \in \RR^N\) stands for the unknown signal in the time domain,
and \(\z \in \CC^P\) contains the (also unknown) coefficients.
As for the linear operators, \(A:\RR^N \rightarrow \CC^P\) is the analysis
(thus $P\geq N$)
and \(D: \CC^P \rightarrow \RR^N\) is the synthesis, while it holds \(D = A^*\).
For computational reasons, we restrict ourselves only to the Parseval tight frames \cite{christensen2008}, i.e.\ \(DD^* = A^*A = \mathit{Id}\), with unitary operators as their special cases.

The problems \eqref{eq:problem_const} can be recast as the sum of two indicator functions:
\begin{equation}
	\argmin_{\x, \z, k} \igg + \il \ \, \text{s.\,t.}
		\begin{cases}
		&\hspace{-1em}\norm{A\x-\z}_2 \leq\epsilon, \\
		&\hspace{-1em}\norm{\x-D\z}_2 \leq\epsilon,
	\end{cases}
\label{eq:problem_unconst}
\end{equation}
where \(\igg\) makes the restored signal lie in \(\Gamma\)
and \(\il\) is a~shorthand notation for \(\iota_{\{\tilde{\z} \hspace{0.1em}|\hspace{0.1em} \|\tilde{\z}\|_0 \leq k\}} (\z)\),
which enforces the sparsity of the coefficients.

The signal is cut into overlapping blocks and windowed prior to processing.
Therefore, in \eqref{eq:problem_unconst}, $\y$ should be understood as one of the signal chunks.
The overall resulting signal is made up by the overlap-add procedure.
As the transformations,
\cite{Kitic2015:Sparsity.cosparsity.declipping} 
uses an overcomplete DFT and IDFT, respectively.

\subsection{A-SPADE}
To solve the analysis variant of \eqref{eq:problem_unconst}, the Augmented Lagrangian is formed according to \eqref{eq:augmented_lagrangian}
and the three ADMM steps according to \eqref{eq:ADMM_unscaled} are constructed:
\begin{subequations}\label{eq:ADMM_ASPADE_const}
\begin{align}
\label{eq:ADMM_ASPADE_const_f}
\hspace{-0.5em}\x^{(i+1)} & = \argmin_\x\|A\x\hspace{-1pt} - \hspace{-1pt}\z^{(i)}\hspace{-1pt} + \hspace{-1pt}\u^{(i)}\|^2_2 \quad \text{s.t.} \   \x \in \Gamma, \\
\label{eq:ADMM_ASPADE_const_g}
\hspace{-0.5em}\z^{(i+1)} & = \argmin_\z\|A\x^{(i+1)}\hspace{-1pt} - \hspace{-1pt}\z\hspace{-1pt} + \hspace{-1pt}\u^{(i)}\|^2_2 \  \text{s.t.} \  \|\z\|_0 \leq k,\\
\label{eq:ADMM_ASPADE_conts_u}
\hspace{-0.5em}\u^{(i+1)} & = \u^{(i)} + A\x^{(i+1)} - \z^{(i+1)}.
\end{align}
\end{subequations}

The report \cite{ZaviskaMokryRajmic2018:SPADE_DetailedStudy} shows in detail that 
the subproblem \eqref{eq:ADMM_ASPADE_const_f} is, in fact, a~projection of \((A^*(\z^{(i)} + \u^{(i)}))\) onto  \(\Gamma\), efficiently implemented as an elementwise mapping in the time domain \cite{Kitic2015:Sparsity.cosparsity.declipping,ZaviskaRajmicPrusaVesely2018:RevisitingSSPADE}.
Furthermore, the solution of \eqref{eq:ADMM_ASPADE_const_g} is obtained by applying the hard-thresholding operator \(\mathcal{H}_k\) to \((A\x^{(i+1)}+\u^{(i)})\),
setting all but \(k\) its largest elements to zero, taking into account the complex conjugate coefficients. 
The
A-SPADE algorithm is finally obtained 
by adding the sparsity relaxation step to the above steps \eqref{eq:ADMM_ASPADE_const}, in which the sparsity of the representation is allowed to increase during iterations.
See Alg.\,\ref{alg:aspade}.

\subsection{S-SPADE original and S-SPADE new}

In the synthesis variant, the situation is different. 
Alg.~\ref{alg:sspade} presents the S-SPADE algorithm from \cite{Kitic2015:Sparsity.cosparsity.declipping}.
Here, the two minimization steps are both carried over \(\z\).
Although this approach is based on the ADMM,
it is explained in \cite{ZaviskaMokryRajmic2018:SPADE_DetailedStudy}
that this algorithm solves a~problem that is different from
\eqref{subeq:synthesis}.
Therefore the original S-SPADE is not really a synthesis counterpart of the A-SPADE.
The report \cite{ZaviskaMokryRajmic2018:SPADE_DetailedStudy} shows that only with unitary operators ($A=D^{-1}$)
do all the three problems coincide.



\makeatletter
\newcommand{\removelatexerror}{\let\@latex@error\@gobble}
\makeatother


\begin{table*}
\noindent
\begin{minipage}{\textwidth}{
\removelatexerror
\small
\begin{minipage}[t]{0.33\textwidth}
\flushleft
\vspace{0pt}
\begin{algorithm*}[H]
\DontPrintSemicolon
	\SetKwInput{KwRequire}{Require}
	\SetKw{KwReturn}{return}
	
	\KwRequire{\(A, \y, \Mr, \Mh, \Ml, s, r, \epsilon\)} \vspace{0.3em}
	\({\hat{\x}^{(0)} = \y, \u^{(0)}=\mathbf{0}, i=0, k=s}\)\;
	\(\bar{\z}^{(i+1)} = \mathcal{H}_k\left(A\hat{\x}^{(i)}+\u^{(i)}\right)\)\;
	\({\hat{\x}^{(i+1)} \hspace{-1pt}=\hspace{-1pt} \argmin_\x\hspace{-1pt}{\|A\x\hspace{-1pt}-\hspace{-1pt}\bar{\z}^{(i+1)}\hspace{-1pt}+\hspace{-1pt}\u^{(i)}\|_2^2}}\;
  \newline\text{\hspace{0.5em}s.t.\hspace{0.5em}}\x \in \Gamma\)\;
	\eIf{\(\|A\hat{\x}^{(i+1)}-\bar{\z}^{(i+1)}\|_2 \leq \epsilon\)}{\textup{terminate}\;}
	{\(\u^{(i+1)}=\u^{(i)}+A\hat{\x}^{(i+1)}-\bar{\z}^{(i+1)}\)\;
	\(i \leftarrow i+1\)\;
	\If{\(i\modulo r = 0\)}{\(k \leftarrow k+s\)\;}
	go to 2\;
	}
	\KwReturn{\(\hat{\x} = \hat{\x}^{(i+1)}\)}
	\caption{\mbox{A-SPADE from \cite{Kitic2015:Sparsity.cosparsity.declipping}}}
	\label{alg:aspade}
\end{algorithm*}
\end{minipage}
\hfill
\begin{minipage}[t]{0.33\textwidth}
\flushright
\vspace{0pt}
\begin{algorithm*}[H]
\DontPrintSemicolon
	\SetKwInput{KwRequire}{Require}
	\SetKw{KwReturn}{return}
	
	\KwRequire{\(D, \y, \Mr, \Mh, \Ml, s, r, \epsilon\)} \vspace{0.3em}
	\({\hat{\z}^{(0)} = D^*\y, \u^{(0)}=\mathbf{0}, i=0, k=s}\)\;
	\(\bar{\z}^{(i+1)} = \mathcal{H}_k\left(\hat{\z}^{(i)}+\u^{(i)}\right)\)\;
	\({\hat{\z}^{(i+1)} = \argmin_\z{\|\z-\bar{\z}^{(i+1)}+\u^{(i)}\|_2^2}}\;
	\newline\text{\hspace{0.5em}s.t.\hspace{0.5em}}	D\z \in \Gamma\hspace{-1em}\)\;
	\eIf{\(\|\hat{\z}^{(i+1)}-\bar{\z}^{(i+1)}\|_2 \leq \epsilon\)}{\textup{terminate}\;}
	{\(\u^{(i+1)}=\u^{(i)}+\hat{\z}^{(i+1)}-\bar{\z}^{(i+1)}\)\;
	\(i \leftarrow i+1\)\;
	\If{\(i\modulo r = 0\)}{\(k \leftarrow k+s\)\;}
	go to 2\;
	}
	\KwReturn{\(\hat{\x} = D\hat{\z}^{(i+1)}\)}
	\caption{S-SPADE from \cite{Kitic2015:Sparsity.cosparsity.declipping}}
	\label{alg:sspade}
\end{algorithm*}
\end{minipage}
\hfill
\begin{minipage}[t]{0.33\textwidth}
\flushright
\vspace{0pt}
\begin{algorithm*}[H]
\DontPrintSemicolon
	\SetKwInput{KwRequire}{Require}
	\SetKw{KwReturn}{return}
	\KwRequire{\(D, \y, \Mr, \Mh, \Ml, s, r, \epsilon\)} \vspace{0.3em}
	\({\hat{\x}^{(0)} = \y, \u^{(0)}=\mathbf{0}, i=0, k=s}\)\;
	\(\bar{\z}^{(i+1)} = \mathcal{H}_k\left(D^*(\hat{\x}^{(i)} - \u^{(i)})\right)\)\;
	\({\hat{\x}^{(i+1)} \hspace{-1pt} = \argmin_\x{ \hspace{-2pt} \|D\bar{\z}^{(i+1)}\hspace{-1pt}-\hspace{-1pt}\x\hspace{-1pt}+\hspace{-1pt}\u^{(i)} \hspace{-1pt} \|_2^2}}\;
	\newline\hspace{0.4em}\text{s.t.}\hspace{0.4em}\x \in \Gamma\)\hspace{-1em}\;
	\eIf{\(\|D\bar{\z}^{(i+1)}-\hat{\x}^{(i+1)}\|_2 \leq \epsilon\)}{\textup{terminate}\;}
	{\(\u^{(i+1)}=\u^{(i)}+D\bar{\z}^{(i+1)}-\hat{\x}^{(i+1)}\)\;
	\(i \leftarrow i+1\)\;
	\If{\(i\modulo r = 0\)}{\(k \leftarrow k+s\)\;}
	go to 2\;
	}
	\KwReturn{\(\hat{\x} = \hat{\x}^{(i+1)}\)}
	\caption{\mbox{S-SPADE proposed}}
	\label{alg:sspade_dr}
\end{algorithm*}

\end{minipage}

}
\end{minipage}
\end{table*}

	Next, we show how the synthesis variant of the SPADE algorithm is derived such that it indeed solves \eqref{subeq:synthesis}.
	First of all, the problem \eqref{eq:problem_unconst} is altered as
	\begin{equation}
		\argmin_{\x,\z, k} \,\igg + \il \ \,\text{s.\,t.}\ \, D\z - \x = 0.
	\label{eq:problem_formulation_synthesis}
	\end{equation}
	Next, the Augmented Lagrangian is formed,
	\begin{equation}
		L_\rho(\x,\y,\z) = \il + \igg + \y^{\top}\hspace{-2pt}(D\z-\x) + \frac{\rho}{2}\|D\z - \x\|^2_2.
	\label{eq:augmented_lagrangian_sspade1}
	\end{equation}
	Using the scaled form, \eqref{eq:augmented_lagrangian_sspade1} appears as 
	\begin{equation}
		L_\rho(\x,\z,\u) = \il + \igg + \frac{\rho}{2}\|D\z - \x + \u\|^2_2 - \frac{\rho}{2}\|\u\|_2^2,
	\label{eq:augmented_lagrangian_sspade2}
	\end{equation}
	leading to the following ADMM steps:
	\begin{subequations}\label{eq:ADMM_SSPADE_const}
		\begin{align}
			\label{eq:ADMM_SSPADE_const_f}
		\hspace{-0.5em}\z^{(i+1)} & \hspace{-2pt} = \argmin_\z\|D\z - \x^{(i)} + \u^{(i)}\|^2_2 \hspace{0.5em} \text{s.t.} \hspace{0.4em} \|\z\|_0 \leq k, \\
			\label{eq:ADMM_SSPADE_const_g}
		\hspace{-0.5em}\x^{(i+1)} & \hspace{-2pt} = \argmin_\x\|D\z^{(i+1)} - \x + \u^{(i)}\|^2_2 \hspace{0.5em} \text{s.t.} \hspace{0.5em} \x \in \Gamma,\\
			\label{eq:ADMM_SSPADE_conts_u}
		\hspace{-0.5em}\u^{(i+1)} & \hspace{-2pt} = \u^{(i)} + D\z^{(i+1)} - \x^{(i+1)}.
		\end{align}
	\end{subequations}

As in Sec.\,\ref{subsec:spade}, adding the sparsity relaxation step and a~termination criterion leads to the final  shape of the proposed S-SPADE algorithm---see Alg.\ \ref{alg:sspade_dr}.

Unlike the original variant of S-SPADE in \cite{Kitic2015:Sparsity.cosparsity.declipping}, where the projection in the frequency domain was required and a~special projection lemma had to be used \cite{ZaviskaRajmicPrusaVesely2018:RevisitingSSPADE}, the projection step \eqref{eq:ADMM_SSPADE_const_g} in the proposed S-SPADE algorithm is a~simple elementwise mapping as is the case of the analysis variant.
	
The solution of the minimization step \eqref{eq:ADMM_SSPADE_const_f} is obtained in Alg.\ \ref{alg:sspade_dr}
by applying the hard-thresholding \(\mathcal{H}_k\) \cite{ZaviskaMokryRajmic2018:SPADE_DetailedStudy}. 
Note that due to the non-orthogonality of \(D\), such a~vector is only an approximate solution to
\eqref{eq:ADMM_SSPADE_const_f}
(in contradiction to A-SPADE where \(\mathcal{H}_k\) solves \eqref{eq:ADMM_ASPADE_const_g} exactly,
cf.\ Alg.\,\ref{alg:aspade}).

The computational cost of the SPADE algorithms is dominated by the signal transformations (i.e.\ the synthesis and analysis). 
All the three algorithms require precisely one synthesis and one analysis per iteration, and  therefore, in theory, the computational complexity of the algorithms is the same.

\section{Experiments and results}

The following experiments were designed to compare the proposed variant of S-SPADE
(denoted $\text{S-SPADE}_{\text{DP}}$ with the index for \qm{done properly})
with the original A-SPADE and with the original S-SPADE from
\cite{Kitic2015:Sparsity.cosparsity.declipping}
($\text{S-SPADE}_{\text{O}}$)
in terms of the quality of restoration and the speed of convergence.

Experiments were performed on five diverse audio files with a 16 kHz sampling rate.
In the preprocessing step, the signals were peak-normalized and then artificially clipped using multiple clipping thresholds \(\tc \in \{0.1,\dots,0.9 \}\).

All audio samples were processed frame-wise, using the 1024-sample-long Hann window with 75\,\% overlap.
The algorithms were implemented in MATLAB 2017a using the \mbox{LTFAT} toolbox \cite{ltfatnote022} for signal synthesis and analysis.
As the signal transformation, the oversampled DFT is used.
The relaxation parameters of all the algorithms were set to \(r=1, s=1 \text{ and } \epsilon = 0.1\).

The restoration quality was evaluated using \(\Delta\)SDR, which expresses the signal-to-distortion improvement in dB, defined as
$
\dsdr{} = \mathrm{SDR}(\x, \hat{\x}) - \mathrm{SDR}(\x, \y),
$
where \(\y\) represents the clipped signal, \(\x\) is the original undistorted signal and \(\hat{\x}\) denotes the restored signal. 
The SDR itself is computed as:
\begin{equation}
\mathrm{SDR}(\u, \vv) = 10\log\frac{\|\u\|_2^2}{\|\u-\vv\|_2^2}\ [\mathrm{dB}].
\label{eq:sdr}
\end{equation}
The advantage of using $\Delta$SDR is that it does not depend on 
whether the SDR is computed on the whole signal or on the clipped samples only
(assuming that the restored signal matches the clipped signal on reliable samples).
	
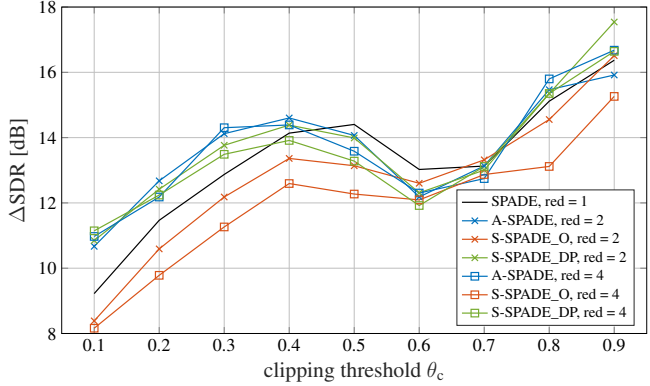
\begin{figure}[t]
\centering
%
%
\definecolor{mycolor1}{rgb}{0.00000,0.44700,0.74100}%
\definecolor{mycolor2}{rgb}{0.85000,0.32500,0.09800}%
\definecolor{mycolor3}{rgb}{0.46600,0.67400,0.18800}%
\begin{tikzpicture}[scale=0.57]

\begin{axis}[%
every axis plot/.append style={thick},
width=5.372in,
height = 3in,
at={(0.462in,0.517in)},
scale only axis,
xmin=0.5,
xmax=9.5,
xtick={0,1,2,3,4,5,6,7,8,9,10},
xticklabels={{},{0.1},{0.2},{0.3},{0.4},{0.5},{0.6},{0.7},{0.8},{0.9},{}},
xticklabel style={font=\large},
xlabel style={font=\Large\color{white!15!black}},
xlabel={$\text{clipping threshold }\theta{}_\text{c}$},
ymin=8,
ymax=18,
yticklabels={{},8,10,12,14,16,18},
ylabel style={font=\Large\color{white!15!black}},
ylabel={${\Delta}\text{SDR [dB]}$},
yticklabel style={font=\large},
axis background/.style={fill=white},
xmajorgrids,
ymajorgrids,
legend style={at={(0.675,0.024)}, anchor=south west, legend cell align=left, align=left, draw=white!15!black}
]
\addplot [color=black]
  table[row sep=crcr]{%
1	9.22243008111303\\
2	11.4674065567651\\
3	12.8716166607615\\
4	14.1374339658972\\
5	14.4036540039474\\
6	13.024411100082\\
7	13.1287045520274\\
8	15.1152183469698\\
9	16.3705909204669\\
};
\addlegendentry{SPADE, red = 1}

\addplot [color=mycolor1, mark=x, mark options={solid, mycolor1}, mark size=3pt]
  table[row sep=crcr]{%
1	10.6694278107238\\
2	12.6761453735851\\
3	14.1212139606546\\
4	14.6022696712219\\
5	14.066543442589\\
6	12.1821749647544\\
7	13.1327184832195\\
8	15.4681647887145\\
9	15.9174927344392\\
};
\addlegendentry{A-SPADE, red = 2}

\addplot [color=mycolor2, mark=x, mark options={solid, mycolor2}, mark size=3pt]
  table[row sep=crcr]{%
1	8.38613759238405\\
2	10.590820286684\\
3	12.1811008442565\\
4	13.3623735232559\\
5	13.1385132029544\\
6	12.6013678199752\\
7	13.3194554007357\\
8	14.5545730920367\\
9	16.5114132583579\\
};
\addlegendentry{S-SPADE\_O, red = 2}

\addplot [color=mycolor3, mark=x, mark options={solid, mycolor3}, mark size=3pt]
  table[row sep=crcr]{%
1	10.8485629834477\\
2	12.418985978878\\
3	13.7652786057716\\
4	14.3834861137123\\
5	13.9894773195784\\
6	12.3091451646643\\
7	13.0053578049331\\
8	15.3487774155301\\
9	17.5414745202682\\
};
\addlegendentry{S-SPADE\_DP, red = 2}

\addplot [color=mycolor1, mark=square, mark options={solid, mycolor1}, mark size=2.5pt]
  table[row sep=crcr]{%
1	10.9724613839466\\
2	12.180415725223\\
3	14.303378818972\\
4	14.3883607684186\\
5	13.5831722017936\\
6	12.3020417260349\\
7	12.744904189442\\
8	15.7980586335679\\
9	16.6783351834808\\
};
\addlegendentry{A-SPADE, red = 4}

\addplot [color=mycolor2, mark=square, mark options={solid, mycolor2}, mark size=2.5pt]
  table[row sep=crcr]{%
1	8.1612663953835\\
2	9.78201541749578\\
3	11.2637014477321\\
4	12.5935384755478\\
5	12.2716010000833\\
6	12.0921938252312\\
7	12.8694911177883\\
8	13.1168480063747\\
9	15.262010053503\\
};
\addlegendentry{S-SPADE\_O, red = 4}

\addplot [color=mycolor3, mark=square, mark options={solid, mycolor3}, mark size=2.5pt]
  table[row sep=crcr]{%
1	11.1451447832039\\
2	12.2404521080999\\
3	13.4889261865367\\
4	13.9139526553656\\
5	13.2809615585512\\
6	11.9199449732678\\
7	13.117335819415\\
8	15.3512835039443\\
9	16.6417334771371\\
};
\addlegendentry{S-SPADE\_DP, red = 4}

\end{axis}
\end{tikzpicture}%
\vspace{-2em}
\caption{Average performance in terms of \(\Delta\)SDR for all three algorithms. Notation \uv{red} denotes redundancy of the DFT. }%
\label{fig:average_dSDR}%
\end{figure}

Fig.\,\ref{fig:average_dSDR} presents the overall $\Delta$SDR results of all SPADE algorithms depending on the clipping threshold \(\tc\).
When no redundancy (orthonormal case) is used, all three algorithms perform equally, which results in the black line. 
With higher redundancies,  both A-SPADE and $\text{S-SPADE}_{\text{DP}}$ 
significantly outperform $\text{S-SPADE}_{\text{O}}$, especially for lower thresholds $\tc$.

\begin{figure}[b!]
\centering
\input{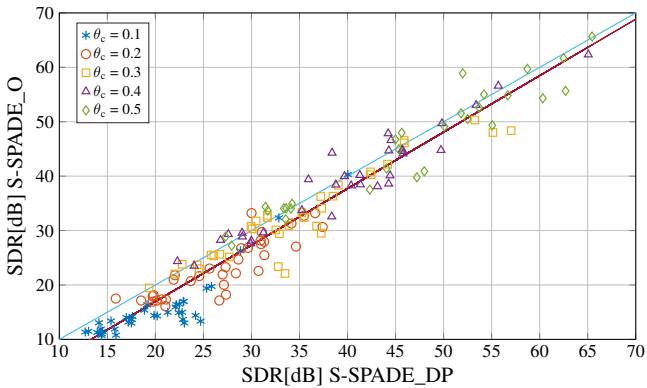}
\vspace{-2em}
\caption{Scatter plot of SDR values for $\text{S-SPADE}_{\text{O}}$ and $\text{S-SPADE}_{\text{DP}}$, computed locally on blocks 2048 samples long.
The blue line is the identity line and the red line represents linear regression.
The results shown are for the signal of acoustic guitar with the twice oversampled DFT (red~=~2).}%
\label{fig:scatter_old_vs_new}%
\end{figure}

\begin{figure}[t]
\centering
\input{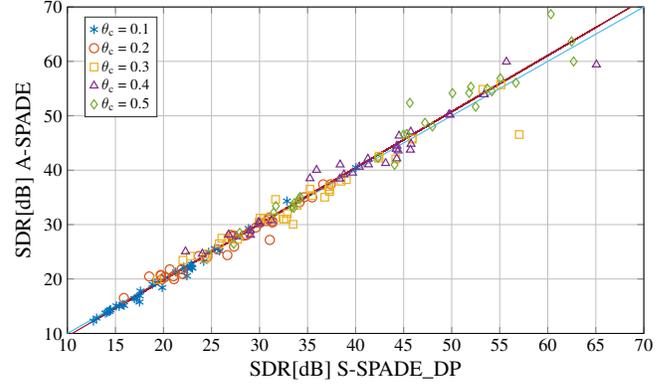}
\vspace{-2em}
\caption{Scatter plot of SDR values for A-SPADE and $\text{S-SPADE}_{\text{DP}}$, computed locally on blocks 2048 samples long.}%
\label{fig:scatter_aspade_vs_sspade}%
\end{figure}

Apart from overall $\Delta$SDR evaluation, a more detailed comparison can be seen on scatter plots in Figs.\,\ref{fig:scatter_old_vs_new} and \ref{fig:scatter_aspade_vs_sspade},
where
$\text{S-SPADE}_{\text{DP}}$
is compared with
$\text{S-SPADE}_{\text{O}}$
and A-SPADE, respectively.
Each mark in the scatter plot corresponds to the SDR value obtained from a~particular 2048-sample-long block. 
For clarity, only results for clipping thresholds from 0.1 to 0.5 are displayed.
Fig.\,\ref{fig:scatter_old_vs_new} displays the linear regression line;
clearly, a~majority of the marks are placed below the blue identity line,
meaning that in most of the time chunks, the $\text{S-SPADE}_{\text{DP}}$ performed better
than $\text{S-SPADE}_{\text{O}}$.
Results from the second scatter plot in Fig.\,\ref{fig:scatter_aspade_vs_sspade} prove an on-par
restoration quality of A-SPADE and $\text{S-SPADE}_{\text{DP}}$, where $\text{S-SPADE}_{\text{DP}}$ performed somewhat better for low SDR and vice versa.

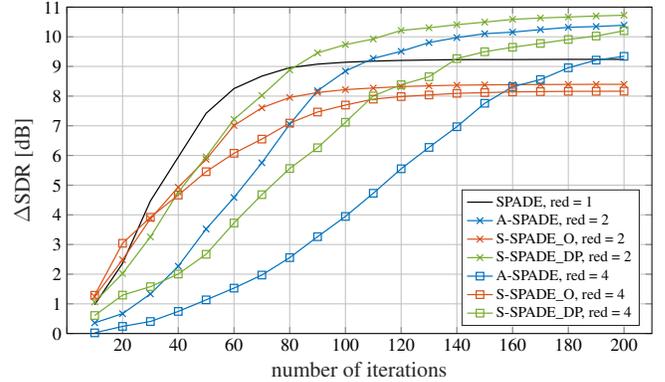
\begin{figure}[h]
\centering
%
%
\definecolor{mycolor1}{rgb}{0.00000,0.44700,0.74100}%
\definecolor{mycolor2}{rgb}{0.85000,0.32500,0.09800}%
\definecolor{mycolor3}{rgb}{0.46600,0.67400,0.18800}%
\begin{tikzpicture}[scale=0.57]

\begin{axis}[%
every axis plot/.append style={thick},
width=5.372in,
height=3in,
at={(0.462in,0.455in)},
scale only axis,
xmin=0,
xmax=21,
xtick={0,2,4,6,8,10,12,14,16,18,20},
xticklabels={{},{20},{40},{60},{80},{100},{120},{140},{160},{180},{200}},
xticklabel style={font=\large},
xlabel style={font=\Large\color{white!15!black}},
xlabel={number of iterations}, 
ymin=0,
ymax=11,
ytick={0,1,2,3,4,5,6,7,8,9,10,11},
ylabel style={font=\Large\color{white!15!black}},
ylabel={${\Delta}\text{SDR [dB]}$},
yticklabels={0,1,2,3,4,5,6,7,8,9,10,11},
yticklabel style={font=\large},
axis background/.style={fill=white},
xmajorgrids,
ymajorgrids,
legend style={at={(0.675,0.024)}, anchor=south west, legend cell align=left, align=left, draw=white!15!black}
]
\addplot [color=black]
  table[row sep=crcr]{%
1	0.971859733628298\\
2	2.36497509407351\\
3	4.46637266348496\\
4	5.94565668723367\\
5	7.42137862163642\\
6	8.25421404601306\\
7	8.67369998861735\\
8	8.95843955089207\\
9	9.07691253166347\\
10	9.14183127589262\\
11	9.17829983953114\\
12	9.2036294604982\\
13	9.2172704995283\\
14	9.2287130035557\\
15	9.23137837997296\\
16	9.23092912077323\\
17	9.23416839724444\\
18	9.23842960403338\\
19	9.23883508793855\\
20	9.23735167356289\\
};
\addlegendentry{SPADE, red = 1}

\addplot [color=mycolor1, mark=x, mark options={solid, mycolor1}, mark size=3pt]
  table[row sep=crcr]{%
1	0.360225620808232\\
2	0.670131414927077\\
3	1.33419165501411\\
4	2.26872722537064\\
5	3.52161953763072\\
6	4.58223668041605\\
7	5.75281044730882\\
8	7.05130943880925\\
9	8.18123127271812\\
10	8.84207389394074\\
11	9.27053834971067\\
12	9.50887033455301\\
13	9.80386364501298\\
14	9.9722261639897\\
15	10.1029743595051\\
16	10.1559798352373\\
17	10.2412346929207\\
18	10.3166668958294\\
19	10.3445222846541\\
20	10.3875321974329\\
};
\addlegendentry{A-SPADE, red = 2}

\addplot [color=mycolor2, mark=x, mark options={solid, mycolor2}, mark size=3pt]
  table[row sep=crcr]{%
1	1.24682146185173\\
2	2.48659624284\\
3	3.87029691653286\\
4	4.92975242143647\\
5	5.86871662866051\\
6	7.00911299773516\\
7	7.60694035313971\\
8	7.95467658788585\\
9	8.12004177540785\\
10	8.21841988461813\\
11	8.27297450630254\\
12	8.31975626507058\\
13	8.34665926160369\\
14	8.36730988368273\\
15	8.37872573099525\\
16	8.38609231153692\\
17	8.39028031748257\\
18	8.39383270495777\\
19	8.39698053345455\\
20	8.39669303684872\\
};
\addlegendentry{S-SPADE\_O, red = 2}

\addplot [color=mycolor3, mark=x, mark options={solid, mycolor3}, mark size=3pt]
  table[row sep=crcr]{%
1	1.06347782706495\\
2	2.02129002484564\\
3	3.25682056033498\\
4	4.76011806073646\\
5	5.95574071098855\\
6	7.21731311458881\\
7	8.0217094392144\\
8	8.88072247091128\\
9	9.45670702394468\\
10	9.73449589571712\\
11	9.92392630955422\\
12	10.2130700434484\\
13	10.3025491348581\\
14	10.4063226218551\\
15	10.4937296871709\\
16	10.5855456943456\\
17	10.6327054592433\\
18	10.6641973899539\\
19	10.7000965889047\\
20	10.724598171199\\
};
\addlegendentry{S-SPADE\_DP, red = 2}

\addplot [color=mycolor1, mark=square, mark options={solid, mycolor1}, mark size=2.5pt]
  table[row sep=crcr]{%
1	0.0237461486188497\\
2	0.240817325449401\\
3	0.405477365704267\\
4	0.744935688416057\\
5	1.13466311237312\\
6	1.5300851890081\\
7	1.97048005440215\\
8	2.55786416126725\\
9	3.26224141469595\\
10	3.94516403719236\\
11	4.72635253623496\\
12	5.55119470650394\\
13	6.27168333474358\\
14	6.96850060199981\\
15	7.75914024171937\\
16	8.31486614242295\\
17	8.55665198689311\\
18	8.95466543477332\\
19	9.21214624891057\\
20	9.34082959956268\\
};
\addlegendentry{A-SPADE, red = 4}

\addplot [color=mycolor2, mark=square, mark options={solid, mycolor2}, mark size=2.5pt]
  table[row sep=crcr]{%
1	1.29044555253549\\
2	3.04059809106158\\
3	3.91817471830909\\
4	4.66486989962554\\
5	5.45352262374462\\
6	6.07395383584027\\
7	6.55149835404471\\
8	7.09010728654827\\
9	7.4637871012083\\
10	7.69833394652398\\
11	7.89783891018555\\
12	7.98350667996965\\
13	8.03955315723652\\
14	8.09313949473873\\
15	8.11777927154315\\
16	8.14047270790492\\
17	8.15070125166878\\
18	8.16020587921897\\
19	8.16233881490489\\
20	8.16520416565034\\
};
\addlegendentry{S-SPADE\_O, red = 4}

\addplot [color=mycolor3, mark=square, mark options={solid, mycolor3}, mark size=2.5pt]
  table[row sep=crcr]{%
1	0.607653289686354\\
2	1.29481781155289\\
3	1.57817072072074\\
4	2.00580639449723\\
5	2.67033267037443\\
6	3.72431376378516\\
7	4.6795788712006\\
8	5.5598215412219\\
9	6.26034298736363\\
10	7.11793475205132\\
11	8.00558302140524\\
12	8.38321746566644\\
13	8.65262315455599\\
14	9.26526949757603\\
15	9.49430164359186\\
16	9.64625086000412\\
17	9.77513757459697\\
18	9.9090788118431\\
19	10.02438693441\\
20	10.2015023457535\\
};
\addlegendentry{S-SPADE\_DP, red = 4}

\end{axis}
\end{tikzpicture}%
\vspace{-2em}
\caption{Average $\Delta$SDR versus the number of iterations.}%
\label{fig:average_dSDR_iterations2}%
\end{figure}

The last experiment compares the SPADE algorithms in terms of the speed of convergence. 
For this purpose, the number of iterations was fixed for each processed block and the $\Delta$SDR was computed from the whole restored signal.
More precisely, the number of iterations varied from 10 to 200
(the termination criterion based on \(\epsilon\) thus does not come into play).
The results are presented in Fig.\,\ref{fig:average_dSDR_iterations2} and they indicate that for redundant operators,
$\text{S-SPADE}_{\text{DP}}$ converges faster than A-SPADE. 
$\text{S-SPADE}_{\text{O}}$
gains the SDR quickly
but for a higher number of iterations it is not able to achieve a~sufficient $\Delta$SDR.

The source codes and sound signals are available at
\href{http://www.utko.feec.vutbr.cz/~rajmic/software/SPADE-DR.zip}{www.utko.feec.vutbr.cz/~rajmic/software/SPADE-DR.zip}.

\section{Conclusion}
A novel algorithm for audio declipping based on the sparse synthesis model was introduced.
Unlike the original S-SPADE, the proposed version really solves the problem formulation \eqref{subeq:synthesis}.
The restoration performance is significantly better than with the original version of S-SPADE and it is comparable with the analysis variant.
The experiments also show that the new S-SPADE converges faster than A-SPADE.

\clearpage
\bibliographystyle{IEEEbib}
\bibliography{literatura}

\newcommand{\noopsort}[1]{} \newcommand{\printfirst}[2]{#1}
  \newcommand{\singleletter}[1]{#1} \newcommand{\switchargs}[2]{#2#1}
\begin{thebibliography}{10}

\bibitem{Tan2003}
Ch.-T.~Tan, B.\,C.\,J.~Moore, and N.~Zacharov,
\newblock ``The effect of nonlinear distortion on the perceived quality of
  music and speech signals,''
\newblock {\em J. Audio Eng. Soc}, vol. 51, no. 11, pp. 1012--1031, 2003.

\bibitem{Adler2011:Declipping}
A.~Adler, V.~Emiya, M.\,G.~Jafari, M.~Elad, R.~Gribonval, and M.\,D.~Plumbley,
\newblock ``A constrained matching pursuit approach to audio declipping,''
\newblock in {\em Acoustics, Speech and Signal Processing (ICASSP), 2011 IEEE
  International Conference on}, 2011, pp. 329--332.

\bibitem{Weinstein2011:DeclippingSparseland}
A.\,J.~Weinstein and M.\,B.~Wakin,
\newblock ``Recovering a clipped signal in sparseland,''
\newblock {\em CoRR}, vol. abs/1110.5063, 2011.

\bibitem{SiedenburgKowalskiDoerfler2014:Audio.declip.social.sparsity}
K.~Siedenburg, M.~Kowalski, and M.~D\"{o}rfler,
\newblock ``Audio declipping with social sparsity,''
\newblock in {\em Acoustics, Speech and Signal Processing (ICASSP), 2014 IEEE
  International Conference on}. 2014, pp.~1577--1581.

\bibitem{Kitic2013:Consistent.iter.hard.thresholding}
S.~Kiti{\'c}, L.~Jacques, N.~Madhu, M.\,P.~Hopwood, A.~Spriet, and
  C.~De~Vleeschouwer,
\newblock ``Consistent iterative hard thresholding for signal declipping,''
\newblock in {\em Acoustics, Speech and Signal Processing (ICASSP), 2013 IEEE
  International Conference on}. May 2013, pp.~5939--5943.

\bibitem{KiticBertinGribonval2014:AudioDeclippingCosparseHardThresholding}
S.~Kiti{\'c}, N.~Bertin, and R.~Gribonval,
\newblock ``Audio declipping by cosparse hard thresholding,''
\newblock in {\em 2nd Traveling Workshop on Interactions between Sparse models
  and Technology}, 2014.

\bibitem{BilenOzerovPerez2015:declipping.via.NMF}
\c{C}.~Bilen, A.~Ozerov, and P.~P\'{e}rez,
\newblock ``Audio declipping via nonnegative matrix factorization,''
\newblock in {\em Applications of Signal Processing to Audio and Acoustics
  (WASPAA), 2015 IEEE Workshop on}, Oct 2015, pp. 1--5.

\bibitem{Kitic2015:Sparsity.cosparsity.declipping}
S.~Kiti{\'c}, N.~Bertin, and R.~Gribonval,
\newblock ``Sparsity and cosparsity for audio declipping: a flexible non-convex
  approach,''
\newblock in {\em {LVA/ICA 2015} -- The 12th International Conference on Latent
  Variable Analysis and Signal Separation}, Liberec, Czech Republic, Aug. 2015.

\bibitem{ZaviskaRajmicPrusaVesely2018:RevisitingSSPADE}
P.~Z{\'a}vi{\v{s}}ka, P.~Rajmic, Z.~Pr{\r{u}}{\v{s}}a, and V.~Vesel{\'y},
\newblock ``Revisiting synthesis model in sparse audio declipper,''
\newblock in {\em {LVA/ICA 2018} -- The 14th International Conference on Latent 
Variable Analysis and Signal Separation}, Guildford, UK, 2018, pp. 429--445, Springer International Publishing.

\bibitem{ZaviskaMokryRajmic2018:SPADE_DetailedStudy}
P.~{Z{\'a}vi{\v s}ka}, O.~{Mokr{\'y}}, and P.~{Rajmic},
\newblock ``{S-SPADE Done Right: Detailed Study of the Sparse Audio Declipper
  Algorithms},''
\newblock techreport, Brno University of Technology, Sept. 2018,
\newblock URL: \url{https://arxiv.org/pdf/1809.09847.pdf}.

\bibitem{Boyd2011ADMM}
S.\,P.~Boyd, N.~Parikh, E.~Chu, B.~Peleato, and J.~Eckstein,
\newblock ``Distributed optimization and statistical learning via the
  alternating direction method of multipliers.,''
\newblock {\em Foundations and Trends in Machine Learning}, vol. 3, no. 1, pp.
  1--122, 2011.

\bibitem{christensen2008}
O.~Christensen,
\newblock {\em Frames and Bases, An Introductory Course},
\newblock Birkh\"{a}user, Boston, 2008.

\bibitem{ltfatnote022}
Z.~Pr\r{u}\v{s}a, P.~S{\o}ndergaard, P.~Balazs, and N.~Holighaus,
\newblock ``{LTFAT: A Matlab/Octave toolbox for sound processing},''
\newblock in {\em Proceedings of the 10th International Symposium on Computer
  Music Multidisciplinary Research (CMMR 2013)}, Marseille, France, October
  2013, {Laboratoire de M\'{e}canique et d'Acoustique}, pp. 299--314, Publications
  of L.M.A.

\end{thebibliography}

\end{document}